\documentclass[conference]{IEEEtran}
\usepackage{verbatim}
\usepackage{lineno,hyperref}
\modulolinenumbers[20]
\usepackage[titlenumbered,ruled]{algorithm2e}
\usepackage{graphicx, subfigure}
\usepackage{color}
\usepackage{booktabs}
\usepackage{longtable}
\usepackage{lipsum}
\usepackage{enumitem}
\usepackage{epsfig}
\usepackage{float}
\usepackage[flushleft]{threeparttable}
\begin{document}

\title{DySign: Dynamic Fingerprinting for the Automatic Detection of Android Malware}

\author{\IEEEauthorblockN{ElMouatez Billah Karbab}
\IEEEauthorblockA{Concordia University\\
e\_karbab@encs.concordia.ca}
\and
\IEEEauthorblockN{Mourad Debbabi}
\IEEEauthorblockA{Concordia University\\
debbabi@encs.concordia.ca}
\and
\IEEEauthorblockN{Saed Alrabaee}
\IEEEauthorblockA{Concordia University\\
s\_alraba@encs.concordia.ca}
\and
\IEEEauthorblockN{Djedjiga Mouheb}
\IEEEauthorblockA{Concordia University\\
d\_mouheb@encs.concordia.ca}
}

\maketitle

\begin{abstract}
The astonishing spread of Android OS, not only in smart phones and tablets but also in IoT devices, makes this operating system a very tempting target for malware threats. Indeed, the latter are expanding at a similar rate. In this respect, malware fingerprints, whether based on cryptographic or fuzzy-hashing, are the first defense line against such attacks. Fuzzy-hashing fingerprints are suitable for capturing malware static features. Moreover, they are more resilient to small changes in the actual static content of malware files. On the other hand, \textit{dynamic analysis} is another technique for malware detection that uses emulation environments to extract behavioral features of Android malware. However, to the best of our knowledge, there is no such fingerprinting technique that leverages dynamic analysis and would act as the first defense against Android malware attacks. In this paper, we address the following question: could we generate effective fingerprints for Android malware through \textit{dynamic analysis}? To this end, we propose {\sf DySign}, a novel technique for fingerprinting Android malware's dynamic behaviors. This is achieved through the generation of a digest from the \textit{dynamic analysis} of a malware sample with respect to existing known malware. It is important to mention that: (i) {\sf DySign} fingerprints are approximates of the observed behaviors during dynamic analysis so as to achieve resiliency to small changes in the behaviors of future malware variants; (ii) Fingerprint computation is agnostic to the analyzed malware sample or family. {\sf DySign} leverages state-of-the-art \textit{Natural Language Processing} (NLP) techniques to generate the aforementioned fingerprints, which are then leveraged to build an enhanced Android malware detection system with family attribution. The evaluation of the proposed system on both real-life malware and benign apps demonstrates a good detection performance with high scalability. 
\end{abstract}

\IEEEpeerreviewmaketitle

\section{Introduction}

The rapid growth in technologies triggers the development and evolvement of mobile devices to enhance both economic and social interactions. Hence, mobile applications (referred to as \emph{apps} henceforth) running on smart devices are gaining ubiquity due to their convenience. For instance, nowadays users purchase products online and in retail stores at their fingertips using such \emph{apps} like Apple pay app. However, the growth of the mobile market \emph{apps} has increased the concerns about apps' security. Android\cite{android_os} is one of the most adopted mobile OS in smart devices, especially in the emerging Internet of Things (\emph{IoT}) world through Brillo \cite{brillokey}, an Android-based IoT system. However, this \emph{IoT} mega-trend makes Android security more crucial than ever before. This is due to the fact that \emph{IoT} devices are everywhere and control important services in cars\cite{android_auto}, TVs\cite{android_tv}, watches\cite{android_wear}, etc. Consequently, this has motivated malware writers to launch attacks against mobile \emph{apps}. These attacks may cause direct financial losses or sensitive data leakages since some \emph{apps} perform monetary transactions using sensitive information such as credit card numbers and passwords. Malware attacks targeting smart devices may also harm IoT devices. This deployment coverage made Android more tempting for cyber-attackers. For example, according to G DATA \cite{GData_2015_stats}, $1,548,129$ and $2,333,777$ new Android malware were discovered in 2014 and 2015, which represents approximately an average of $4,250$ and $6,400$ new malware per day respectively. Furthermore, about $53$\% of malware are SMS Trojans designed to steal funds and personal information from Android-based mobile devices \cite{Canfora:2015}. 

In this context, it is a desideratum to develop a scalable, efficient, and accurate framework that tackles two distinct problems: (i) Malware detection - distinguishing malicious from benign applications, and (ii) malware family attribution - assigning malware samples to known families. 

%While the former immediately benefits users, the latter is a crucial part for forensic analysis and threat assessment.

\textbf{Problem Statement:} In the literature, malware analysis may be categorized into static and dynamic analyses. In static analysis, fingerprints are the first defense line against malware attacks. Two common static analysis techniques are used for fingerprinting Android malware, namely, cryptographic and fuzzy/approximate hashing techniques. Using cryptographic hashes may be easily defeated by the tiniest change in malware Android packaging (APK). The \textit{Fuzzy/approximate hashing} technique is more resilient to small changes. Moreover, it has the possibility of detecting malware variants produced by APK repacking. On the other hand,  dynamic analysis misses a fuzzy fingerprint, similarly to APK file fuzzy fingerprint, that could effectively capture Android malware run-time behaviors instead of APK static content. Dynamic analysis is commonly used to obtain dynamic features that are fed to a classifier to detect Android malware or cluster them according to their families. However, this dynamic analysis process suffers from the following main drawbacks: i) The detection needs an intermediate, which is the learning model, between the dynamic analysis of malicious Android APK and the new app to check its similarity; ii) in most cases, the extracted dynamic features are driven by the malware dataset. Accordingly, we choose features that give the most accurate model. Although these features could fingerprint the malware family of the dataset, it is hard to predict if extracting the same features from other Android malicious apps could fingerprint them. As such, malware or family-agnostic features are needed in order to have a resilient fingerprinting technique; iii) directly using dynamic analysis output to compare between malware dynamic behaviors lacks portability due to the inconsistent sizes of the output. Moreover, there is no defined approach for similarity computation.

The aforementioned drawbacks induce the need for a dynamic analysis based fingerprint with a fixed size to achieve portability and  compute similarity between malware dynamic behaviors. Such fingerprint should be agnostic to the malware sample or family. Hence, the fingerprint extraction approach needs to be general enough to cover most of the essential information of the dynamic analysis output in most Android malware. In addition, it needs to be scalable to compute a digest relatively to the known malware and achieve a fast detection decision. To the best of our knowledge, there is no such a fingerprinting technique that abstracts the dynamic analysis in one digest for the purpose of Android malware detection.

\textbf{Approach Overview:} In this paper, we propose a novel fingerprinting approach, namely \textsf{DySign}, which aims at generating a signature that is based on the dynamic analysis of Android malware apps. In particular, the proposed approach aims at achieving the following properties: (i) \textsf{DySign} fingerprints are approximates of the observed behaviors during dynamic analysis so as to achieve resiliency to small changes in the behaviors of future malware variants; (ii) fingerprint computation is agnostic to the analyzed malware sample or family. We choose these properties since they allow our proposed approach \textsf{DynSign} to efficiently detect malware variants and other samples of the same family efficiently. The key idea of \textsf{DySign} lies in the fact that Android malicious apps, such as SMS Trojans, tend to have similar overall dynamic behaviors, which are distinguishable from the behaviors of benign apps. In addition, apps targeted by a given malware tend to share  similar behaviors than apps that are targeted by different malware families. In a sandboxing environment, malware runtime behaviors are translated into an analysis report. Therefore, malicious apps with similar behaviors would produce similar analysis reports. In the context of \textsf{DySign}, we leverage the output of Android malware dynamic analysis using \textit{sandboxing} environments to generate \textit{relative} fingerprints from the known Android malware apps analysis reports. More precisely, \textsf{DySign} leverages state-of-the-art Natural Language Processing (NLP) techniques to produce the aforementioned fingerprints using the \textit{bag of words} model in the \textsf{DySign} generation from the analysis reports. Considering the latter as a word makes \textsf{DySign} completely agnostic to the malware sample or family. Furthermore, we leverage \textsf{DySign} to build an enhanced Android malware detection system with family attribution. \textsf{DySign} is evaluated on both real-life malware and benign apps and the obtained results demonstrate  high detection and attribution performances.

\textbf{Contributions:} In summary, this paper makes the following contributions:

\begin{enumerate}
\item We introduce \textsf{DySign}, a novel fingerprinting system for automatically generating dynamic fingerprints for dynamic analysis of Android malware apps.

\item We leverage state-of-the-art of Natural Language Processing (NLP) techniques in order to propose an approach that is resilient to change in the dynamic behaviors of Android malicious apps.
\item We conduct a large-scale evaluation of \textsf{DySign} using $8,639$  malicious and benign apps. Our evaluation demonstrates that \textsf{DySign} achieves a good detection performance with high scalability.
\end{enumerate}

The remainder of this paper is organized as follows: Section \ref{sec:scenarios} presents some usage scenarios of \textsf{DySign}. Section \ref{sec:background} gives a light background on Android OS. Section \ref{sec:methodology} details our methodology. We evaluate \textsf{DySign} in Section \ref{sec:evaluation}. In Section \ref{sec:related_work}, we discuss the related work. In Section \ref{sec:conclusion}, we provide some concluding remarks on this research together with a discussion of future research.

\section{Usage Scenarios} \label{sec:scenarios}

The main aim of \textsf{DySign} is to generate an approximate fingerprint from the dynamic analysis of malicious apps. The fingerprint is generated with respect to a database of known apps analysis. Our main concern after accuracy is scalability of the fingerprinting since \textsf{DySign} is intended to be the first fingerprint's defense line, along with static file fuzzy fingerprint, to tackle the overwhelming volume of malicious apps on a daily basis. \textsf{DySign} has two main usage scenarios: i) \textit{Mobile OS monitoring}: In this scenario, we have a set of installed apps that run in a given smart device (the number of apps could be fixed in the case of an IoT device since it is mono-task with a deterministic goal). Having a runtime report database of the these apps would help \textsf{DySign} to periodically fingerprint the behaviors of these apps in order to check for the existence of abnormal behaviors. In this scenario, \textsf{DySign} could raise an exception of behavior change after a suspicious update or a hack; ii) \textit{Cloud service analyzer}: In this scenario, \textsf{DySign} is used as a core of cloud checking service of the received analysis reports, either automatically or by user submission, from the Android device of suspicious apps. The goal is to match the runtime analysis against malicious apps. These scenarios are general applications of \textsf{DySign}. However, we believe that it can be extended to many other usages due to the simplicity and scalability of \textsf{DySign}.

\section{Background} \label{sec:background}
%This section first describes Android architecture. Then, it introduces Android apps. Finally, a brief overview about Android threats is provided.

\subsection{Android Architecture}
Android has been settled by Android Open Source Project (AOSP) team, maintained by Google and supported by the Open Handset Alliance (OHA)~\cite{oha_handset}. It encompasses the Original Equipment Manufacturers (OEMs), chip-makers, carriers and application developers. Android apps are written in Java. However, the native code and shared libraries are generally developed in C/C++~\cite{android_ndk}. The typical Android architecture consists of Linux kernel, which is designed for an embedded environment consisting of limited resources. 
%Android supports two Instruction Set Architectures: 1) ARM, prevalent on smartphones and tablets; 2) x86, prevalent among mobile internet devices. 
On top of Linux kernel, the native libraries developed in C/C++ support high-performance third-party reusable shared libraries. Moreover, Android apps written in Java are translated into Dalvik bytecode. It is specifically optimized for resource-constrained mobile OS platforms. 
\subsection{Android Threats}
Once the app is installed, it may create undesirable consequences for the device security. Following are some examples of malicious activities that have been reported: i) Personal-information leakage occurs when users give dangerous permissions to malicious apps and unknowingly allow access to sensitive data and its exfiltration without user knowledge or consent; ii) malicious apps can also spy on the users by monitoring their voice calls, SMS/MMS, recording audio/video without user knowledge or consent; iii) compromising the device to act as a bot and remotely control it through a server by sending various commands to perform malicious activities.

%===============================================

\section{DYSIGN Methodology} \label{sec:methodology}
%In this section, we present the architecture overview of DySign along with the different phases of the proposed system.

\begin{figure*}[!htb]
  \centering
      \includegraphics[width=0.7\textwidth]{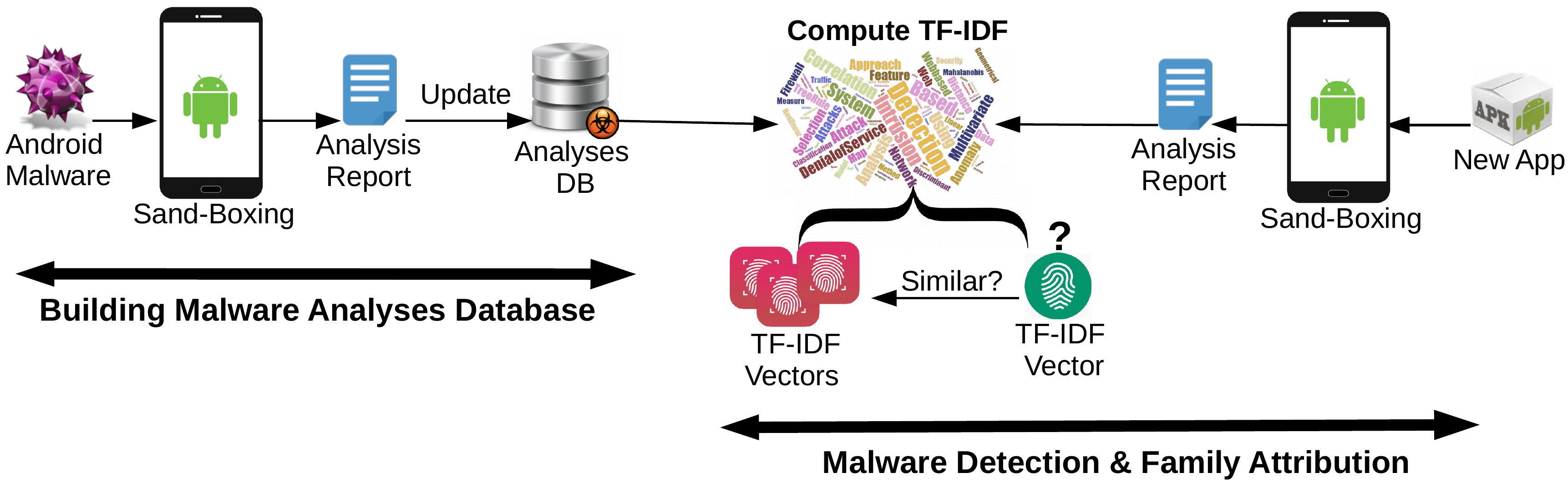}
  \caption{Approach Overview}
  \label{fig:approach_overview}
\end{figure*}

\subsection{Fingerprint Computation}
Our ultimate goal is to automatically fingerprint Android malware based on dynamic analysis. To this end, we use natural language processing techniques, where we consider the output of the dynamic analysis as a plaintext file and model it as a \textit{bag of words}. The latter treats the text document as a set of words separated by predefined delimiters such as \emph{spaces} and \emph{curly-brackets}. Given a set of analysis report \emph{bag of words}, we compute a \emph{relative fingerprint} for each report based on the word frequency in one document and the rest. In other words, we distinguish between the reports by giving a high weight to the words with a high frequency in the given report and low frequency in the others. The result is a vector of words' weights for each analysis report. To compute \textsf{DySign}'s vector, we leverage the so-called \emph{Term Frequency-Inverse Document Frequency} \emph{tf-idf} \cite{itidf_wiki}, a well-known technique adopted in the fields of \emph{information retrieval} and \emph{natural language processing}. The latter computes vectors of inputted text documents by considering both the frequency in the individual documents and in the whole set. Let $D=\{d_1, d_2,\ldots, d_n\}$ be a set of text documents,  where $n$ is the number of documents, and let $d=\{w_1, w_2,\ldots, w_m\}$ be a document, where $m$ is the number of words in $d$. The \emph{tf-idf} of a word $w$ and document $d$ is the product of \emph{term frequency} of $w$ in $d$ and the \emph{inverse document frequency} of $w$, as shown in Formula \ref{equ:tfidf}. The \emph{term frequency} (Formula \ref{equ:tf}) is the occurrence number of $w$ in $d$. Finally, the \emph{inverse document frequency}  of $w$ (Formula \ref{equ:idf}) represents the number of documents $n$ divided by the number of documents that contain $w$ in the logarithmic form. The computation of \emph{tf-idf} is very scalable, which suites our needs (Section \ref{sec:evaluation}).

\begin{equation}
\mbox{\it tf-idf}(w,d) = \mbox{\it tf}(w,d) \times \mbox{\it idf}(w)
\label{equ:tfidf}
\end{equation}

\begin{equation}
\mbox{\it tf}(w,d) = |w \in d, ~d=\{w_1, w_2, ... w_n\}: w = w_i|
\label{equ:tf}
\end{equation}

\begin{equation}
\mbox{\it idf}(w) = log{\frac{|D|}{1 + |d: w \in d|}}
\label{equ:idf}
\end{equation}

The result of \emph{tf-idf} is a set of vectors $V=\{v_1, v_2, \ldots, v_n\}$ (\textsf{DySign} fingerprints) of word weights for each document $d \in D$. Computing the similarity using \textit{DySign} is straightforward using the \textit{cosine similarity} as shown in Formula \ref{equ:cos}.

\begin{equation}
\mbox{cosine-similarity}(v_1, v_2) = \mbox{cos}(\theta) = \frac{v_1 \cdot v_2}{||v_1||||v_2||}
\label{equ:cos}
\end{equation}

\emph{How \textsf{DySign} could be used for Android malware detection and family attribution?} We answer this question through an illustrating example, in which we compute \textsf{DySign} fingerprints from the analysis reports of malware samples from Drebin malware dataset \cite{arp2014drebin,spreitzenbarth2013mobile} along with benign apps downloaded from Google Play \cite{google_play}.  The example is summarized in Table \ref{tab:insight_detection} and Table \ref{tab:insight_attribution}. How \textsf{DySign} is used for malware detection is illustrated in Table \ref{tab:insight_detection}, where we compute the similarity between malware analysis reports and benign ones. This example shows the potential of \textsf{DySign} in distinguishing between malware and benign apps. 

\begin{table}[!h]
\centering
\begin{scriptsize}
\begin{threeparttable}
\begin{tabular}{|c||c|c|c|}
    \hline \hline
     \#&  \textit{App1}     & \textit{App2}        & \textit{TFIDF Cosine}  \\ \hline\hline
    1 & 00453ca8 (FakeInst)\tnote{1} & com.BigBawb.coin.apk &  0.19 		\\ \hline
    2 & 00453ca8 (FakeInst) & com.interestcalculator.apk  & 0.21	\\ \hline 
    3 & 21262a59 (FakeInst) & com.sleggi.MiFreetime.apk  & 0.16 	\\ \hline 
    4 & 00453ca8 (FakeInst) & 21262a59 (FakeInst)  & 0.42 			\\ \hline
    5 & 00453ca8 (FakeInst) & com.sleggi.MiFreetime.apk  & 0.27 	\\ \hline\hline 
\end{tabular}
\begin{tablenotes}
\item[1] First 8 characters from malware hash and its malware family.
\end{tablenotes}
\end{threeparttable}
\caption{Insight of \textsf{DySign} Android Malware Detection} 
\label{tab:insight_detection}
\end{scriptsize}
\end{table}

As shown in Table \ref{tab:insight_attribution},  \textsf{DySign} could be used to segregate between Android malware families by requiring a higher similarity between the fingerprints of the same malware family. Based on these insights, we generalize and build a system on top of \textsf{DySign} for Android malware detection and family attribution.

\begin{table}[!h]
\centering
\begin{scriptsize}
\begin{threeparttable}
\begin{tabular}{|c||c|c|c|}
    \hline \hline
     \#&  \textit{Malware1} & \textit{Malware2}		  & \textit{TFIDF Cosine}   \\ \hline\hline
    1 & 090b5be2 (Plankton)\tnote{2} & bedf51a5 (DroidKungFu)  &  0.56 \\ \hline
    2 & 149bde78 (Plankton) & bedf51a5 (DroidKungFu)  &  0.46    		\\ \hline 
    3 & 090b5be2 (Plankton) & 149bde78 (Plankton)     &  0.71    		\\ \hline\hline 
\end{tabular}
\begin{tablenotes}
\item[2] First 8 characters from malware hash and its malware family.
\end{tablenotes}
\end{threeparttable}
\caption{Insight of \textsf{DySign} Android Malware Family Attribution } 
\label{tab:insight_attribution}
\end{scriptsize}
\end{table}

\emph{How \textsf{DySign} is agnostic to malware samples and families?} \textsf{DySign} is agnostic by design since no features are extracted specifically for a given malware family or sample. In other words, \textsf{DySign} considers the analysis report as a bag of words. It only considers the frequency of the word in a document relatively to the other ones. This ensures that the extracted \textsf{DySign} information is broad enough to cover most malware samples and without relying on specific features.

\subsection{Architecture Overview}
In this section, we present the architecture of \textsf{DySign} framework for Android malware detection, built on top of \textsf{DySign}'s fingerprint. There are two main processes in \textsf{DySign} framework, as depicted in Section \ref{fig:approach_overview}. i) The first process is building the analysis report database. The initial phase of this process consists of a bulk sandboxing and reports insertion into a database of known Android malware (Algorithm \ref{alg:init_db}). Afterwards, the process proceeds as a continuous task of updating the report's database with new apps (Algorithm. \ref{alg:update_db}).

\begin{algorithm}
  \footnotesize {
    \SetKwFunction{SB}{SandBoxing}
    \SetKwFunction{TFIDF}{TFIDF}
    \SetKwFunction{COS}{Similarity}
    \SetKwFunction{DC}{getDecision}
    \SetKwFunction{FM}{getAndroidFamily}
    \SetKwFunction{GB}{getWordBag}
    \SetKwFunction{SD}{SaveDatabase}
    \SetKwFunction{UP}{LunchUpdateProcess}
    
    \SetKwInOut{Input}{Input}\SetKwInOut{Output}{Output}

    \Input{$MalDataset$: APK Files of Known Malware \newline
    		$BenDataset$: APK Files of Some Benign Apps}
    \BlankLine

    \Begin{
      \ForEach {$Apk \in MalDataset$}{
       	$Report \gets$ \SB{$Apk$}\;
       	$WordBag \gets$ \GB{$Report$}\;
		\SD{$WordBag$}\;
      }

      \ForEach {$Apk \in BenDataset$}{
       	$Report \gets$ \SB{$Apk$}\;
       	$WordBag \gets$ \GB{$Report$}\;
		\SD{$WordBag$}\;
      } 
      
      \UP()
    }
  }
  
\caption{First Setup of Analysis Report Database}
\label{alg:init_db}
\end{algorithm}

\begin{algorithm}
  \footnotesize {
    \SetKwFunction{SB}{SandBoxing}
    \SetKwFunction{TFIDF}{TFIDF}
    \SetKwFunction{COS}{Similarity}
    \SetKwFunction{DC}{getDecision}
    \SetKwFunction{FM}{getAndroidFamily}
    \SetKwFunction{GB}{getWordBag}
    \SetKwFunction{SD}{SaveDatabase}
        
    \SetKwInOut{Input}{Input}\SetKwInOut{Output}{Output}

    \Input{$NewUpdateApp$: Update App File (APK)}
    \BlankLine

    \Begin{
	  \While{True}{
	  	\If{$\exists~NewUpdateApp$}{
      		$NewReport \gets$ \SB{$NewUpdateApp$}\;
      		$WordBag \gets$ \GB{$NewReport$}\;
			\SD{$WordBag$}\; 
      	}     	  
	  }	  
    }
  }
\caption{Updating Analysis Report Database}
\label{alg:update_db}
\end{algorithm}

ii) The second process is the detection process, in which we check the runtime behaviors of newly received apps against known malware behaviors.  First, the new app is executed in a sandboxing environment during a time $T$ to get the analysis report. The latter will be used along with the database reports to compute the \textsf{DySign} fingerprint using \textit{tf-idf}. Finally, we compute the similarity between the \textsf{DySign} fingerprint of the new app and the existing fingerprints to identify whether it is malicious or not and its  family in case it is malicious. The complete \textsf{DySign}  process is presented in Algorithm \ref{alg:dysign_detection_process}. Using \textsf{DySing}'s fingerprint, we do not only detect malware but also attribute the unknown samples to their Android malware families. Further, we can also ascribe a family to the unknown samples if we already have samples of this family in \textsf{DySign}'s dynamic analysis database. Algorithm~\ref{alg:dysign_detection_process} describes the process of generating a dynamic fingerprint. 

\begin{algorithm}
  \footnotesize {
    \SetKwFunction{SB}{SandBoxing}
    \SetKwFunction{TFIDF}{TFIDF}
    \SetKwFunction{COS}{Similarity}
    \SetKwFunction{DC}{getDecision}
    \SetKwFunction{FM}{getAndroidFamily}
    \SetKwFunction{GB}{getWordBag}
    
    \SetKwInOut{Input}{Input}\SetKwInOut{Output}{Output}

    \Input{$Database$: Analysis Reports Of \textsf{DySign} Database \newline
    		$NewApp$: New App File (APK)}
    \Output{$Decision$:  \{Bengin or Malicious\} \newline
    		$Family$: Android Malware Family}
     \BlankLine

    \Begin{
      $NewReport \gets$ \SB{$NewApp$}\;
      $dbVectors, NewVector \gets$ \TFIDF{${dbReports}, {NewReport}$}\;

      \BlankLine
      $MaxSim \gets 0$\;
      $Decision \gets$ \texttt{Benign}\;
      $Family \gets \emptyset$ \;

      \ForEach {$Vec \in dbVectors$}{
        $Sim \gets$ \COS{${Vec}, {NewVector}$}\;
        \If {$Sim > MaxSim$}{
          $MaxSim \gets Sim$\;
          $Decision \gets$ \DC{$Vec$}\;
          $Family \gets $ \FM{$Vec$}\;
        }
      }

      \Return ${Decision},{Family}$\;
    }
  }
\caption{DySign Framework Detection Process}
\label{alg:dysign_detection_process}
\end{algorithm}

A cornerstone in \textsf{DySign} framework is the sandboxing system, which heavily influences  the produced analysis reports. We use \emph{DroidBox} \cite{droidbox_github}, a well-established sandboxing environment based on the Android software emulator \cite{android_emulator} provided by Google Android SDK \cite{android_sdk}. Running the app may not lead to a sufficient coverage of the executed app. As such, to simulate the user interaction with the apps, we leverage \textit{MonkeyRunner} \cite{monkeyrunner}, which produces random UI actions aiming for a broader execution coverage. However, this makes the app execution non-deterministic since \textit{MonkeyRunner} generates random actions. Therefore, this yields different analysis reports for every execution, where the accuracy of the results may vary. To tackle this issue, we run the app in a sandboxing environment for a long time $T$ in order to assure the maximum of information in the resulting report. However, a long time $T$ could lead to execution bottleneck since \emph{DroidBox} can only handle  one app at a time. In this context, executing the dataset apps in a sandboxing environment during the initial setup of a \textit{reports database} is a computation bottleneck in \textsf{DySign}. This is because of the defined time $T$, during which the app needs to run in order to get the analysis report. Hence, the initialization phase could take a very long time (may reach few days). To overcome this challenge, we develop a multi-worker sandboxing environment to exploit the maximum available resources and boost the initialization setup.% as depicted in Figure \ref{fig:sandboxing}.
%\begin{figure}[!htb]
%  \centering
%      \includegraphics[width=0.5\textwidth]{sendbox}
%  \caption{Boosting sandboxing with Multi-Instance System}
%  \label{fig:sandboxing}
%\end{figure}
Another problem is the similarity computation, which could be a bottleneck for the \textsf{DySign} framework and could lead to inefficient matching against new unknown apps. To address this issue, we resort to LSH K-Nearest Neighbor (KNN) \cite{lshforest05bawa}.  Similarity computation needs to be conducted in an efficient way that is much faster than the brute-force computation. To this end, we leverage \textit{Locality Sensitive Hashing} (LSH) techniques, and more precisely \textit{LSH Forest} \cite{lshforest05bawa}, a tunable high-performance algorithm for similarity computation. The key idea behind \textit{LSH Forest} is that similar items hashed using LSH are most likely to be in the same bucket (collide) and dissimilar items in different ones as we will explain it in  the next Section.

\subsection{Locality-Sensitive Hashing}
Our system employs Locality-Sensitive Hashing (LSH) for feature reduction~\cite{dasgupta2011fast, lshforest05bawa}. The main idea in LSH is to define a hash function {\tt h} such that {\tt h(s1) = h(s2)} if the two sets of chains $s_{1}$ and $s_{2}$ are similar~\cite{andoni2006near}. The hash is calculated over all sets of traces, and only those with similar hash values are clustered (hashed) to the same bucket. In the case of similarity, similar traces will be hashed to the same bucket. In our case, we assume that most dissimilar pairs will never hash to the same bucket, and therefore will never be checked. Once all traces have been hashed to a corresponding bucket, any bucket containing more than one hash value is identified and a list of candidate traces is extracted. Finally, similarity analysis is performed to rank the candidate pairs obtained from the previous steps. To create the signature from traces, we must use one of the hash function pairs. We choose $minhash$ due to its efficiency.  When using $minhash$ (with $N$ unique hash functions) as signatures to represent the register chains, LSH can be used by splitting the minhash values into a signature matrix with $b$ bands consisting of $r$ rows each. Depending on the number of used bands, the number of minhash values for a given band will be the number of minhashes divided by the number of bands $( N / \# bands)$. The number of rows will be equal to the number of register chain minhash signatures. Finally, for each band $b$, the minhash values (the portion of one column within that band) are hashed to one bucket of a larger number of buckets. 

%===================================================

\section{Experimental Results} \label{sec:evaluation}
In this section, we present the evaluation results of our proposed system. The implementation  subsection shows the setup of our experiments.  To evaluate the performance of malware detection  using \textsf{DySign}, we use a mixed dataset, i.e., malware and benign apps. As for the evaluation of the attribution performance, we use a malware-only dataset.

\subsection{Implementation} \label{sec:implementation}
For modularity purposes, \textsf{DySign} is implemented using separate Python scripts, which altogether form our analytical system. The scripts are used for parsing, cleaning, and tf-idf computation. We develop a multi-sandboxing system on top of \textit{DroidBox} to be able to execute multiple Android apps simultaneously by leveraging the multicore CPUs to have numerous instances of \textit{DroidBox}. We also use \textit{MonkeyRunner} to simulate UI interaction with the user. SQLite has been used to store the features due to its efficiency and ease of use. 

\subsection{DataSet}
The first step towards the evaluation of \textsf{DySign} is to select appropriate datasets that can be utilized for Android malware fingerprinting. Obtaining representative datasets is a fundamental challenge, and there is certainly a strong need for standard ones. Hence, the utilized dataset consists of: i) \textit{malware-only} dataset using the well-known Drebin dataset \cite{arp2014drebin, spreitzenbarth2013mobile}, and ii) \textit{mixed} dataset using Drebin dataset along with benign apps downloaded from Google Play \cite{google_play}. Statistics about the dataset are presented in Table ~\ref{tab:dataset_describe} and Table \ref{tab:dataset_family_number}. In Table\ref{tab:dataset_describe}, we use a subset of $3,414$ Android malware samples, from Drebin dataset, distributed on $8$ families \ref{tab:dataset_family_number}. From this dataset, we exclude all malware families with only few samples due to the high skewness of the dataset. This would  prevent having, for instance, a family with $800$ samples and other families with only 1, 2, or even 20 samples.

\begin{table}[!htb]
\centering
\begin{tabular}{|c||c|c|}
    \hline \hline
      & \textit{Drebin Dataset} & \textit{Drebin Mixed With Benign}  \\\hline\hline
    Total Size  & 3414  & 8639  \\ \hline
    Malware 	& 3414  & 3414  \\ \hline
    Benign  	& / 	& 5225  \\ \hline \hline
\end{tabular}
\caption{Android Dataset Description} 
\label{tab:dataset_describe}
\end{table}

\begin{table}[!htb]
\centering
\begin{scriptsize}
\begin{tabular}{|l||l|c|}
\toprule
{} & Malawre Family & Number of Samples \\ \hline
\midrule 
0 &  FakeInstaller & 866 \\ \hline 
1 &    DroidKungFu & 611\\ \hline
2 &         Opfake & 566\\ \hline 
3 &       Plankton & 515\\ \hline 
4 &      GinMaster & 314\\ \hline 
5 &     BaseBridge & 295\\ \hline 
6 &       Iconosys & 127\\ \hline 
7 &        FakeDoc & 120\\ \hline 
8 &    Benign Apps & 5225 \\ \hline \bottomrule
\end{tabular}
\end{scriptsize}
\caption{Dataset Description By Malware Family} 
\label{tab:dataset_family_number}
\end{table}

\subsection{Results} \label{sec:results}
To evaluate our approach using the previous datasets, we split the training data into ten sets, reserving one set as a testing set and using nine sets as training sets.  We repeat this process numerous times. We use \emph {precision} (P) and \emph {recall} (R):

\begin{equation}
P =\frac{TP}{ {TP + FP}}, ~R =\frac{TP}{ {TP + FN}}, ~F1 = 2 \times \frac{P \times R}{P + R}
\label{equ:precision_recall_fscore}
\end{equation}

\textbf{Detection Performance:} Since the application domain targeted by \emph{DySign} is much more sensitive to false-positives than false-negatives, we employ the F-measure, where the results of $F_{1}$ measure are summarized in Table~\ref{tab:accuracy_results}. We use two types of datasets: (i) The mixed dataset, used for detection performance assessment, and (ii) the malware-only dataset, used to assess \textsf{DySign}'s family attribution, as shown in Table \ref{tab:accuracy_results}. The obtained results show that our approach achieves good detection and attribution performance in short time.

\begin{table}[!h]
\centering
\begin{tabular}{|c||c|c|c|c|}
    \hline \hline
     &   \textit{F1-Score} & \textit{Precision} & \textit{Recall} & \textit{Time} \\
    \hline\hline
    \textbf{Mixed (Detection)} & \textbf{85\%} & \textbf{94\%} & \textbf{78\%} & \textbf{4min 45s}		\\\hline 
    \textbf{Drebin (Attribution) } & \textbf{80\%} & \textbf{82\%} &  \textbf{79\%} & \textbf{2min 20s} \\\hline\hline
\end{tabular}
\caption{Detection and Attribution Performance of \textsf{DySign} } 
\label{tab:accuracy_results}
\end{table}

\begin{scriptsize}
\begin{figure}[ht!]
     \begin{center}
        \subfigure[\# Malware Samples]{%
            \label{fig:number_conf_matrix}
            \includegraphics[width=0.24\textwidth]{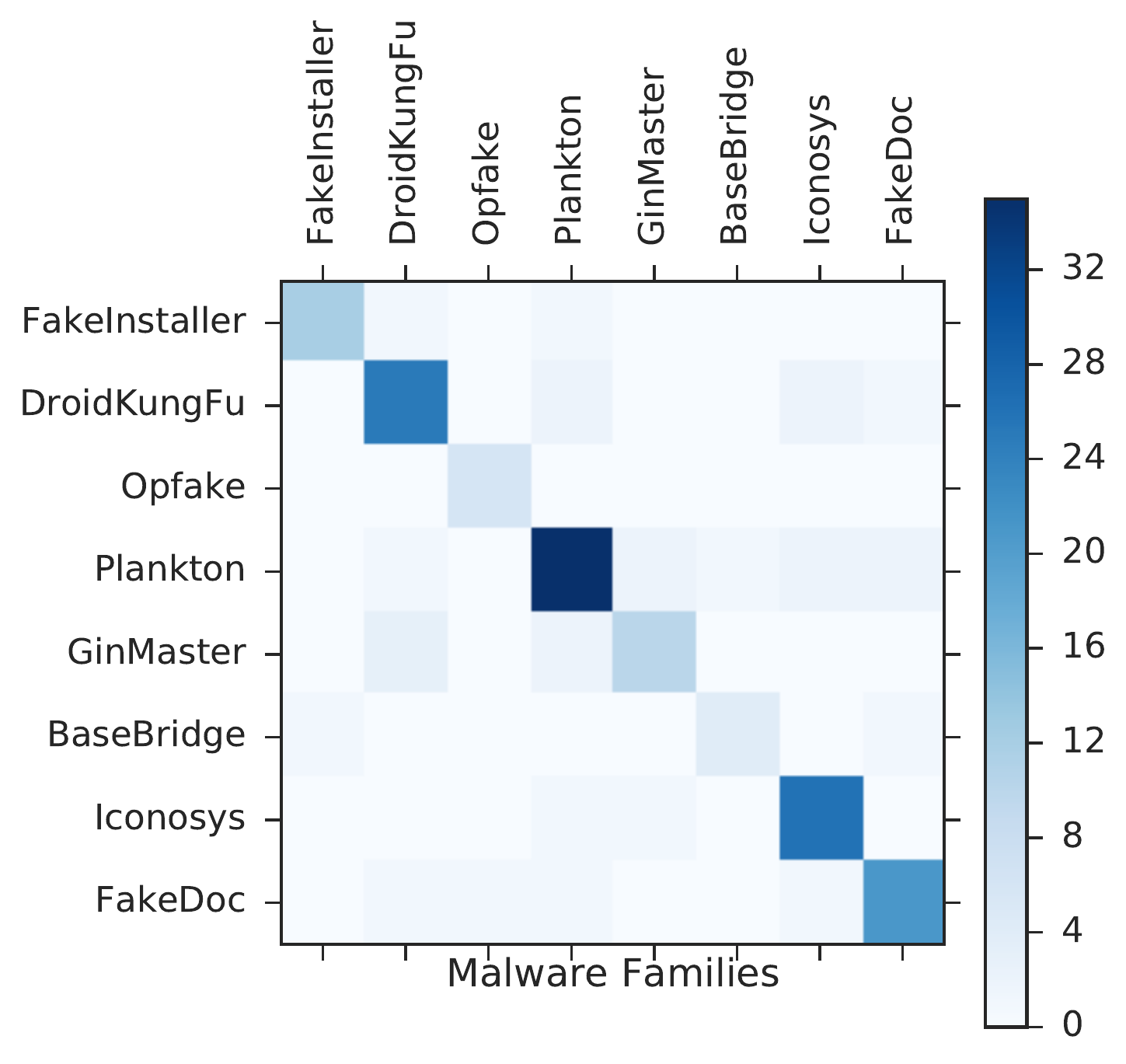}
        }
        \subfigure[Log10(\#)] {%
           \label{fig:log_conf_matrix}
           \includegraphics[width=0.2\textwidth]{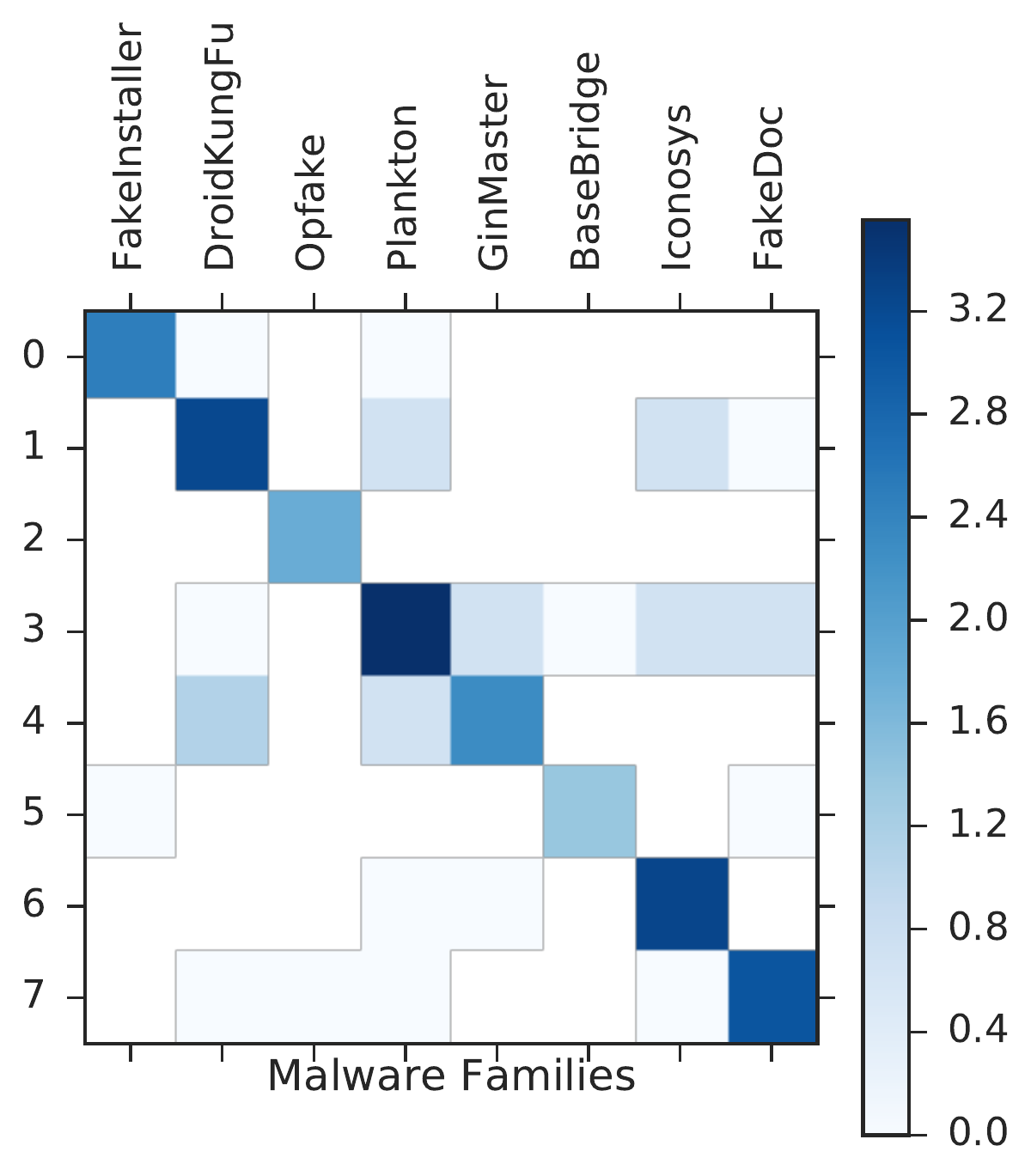}
        }
    \end{center}
    \caption{
        \textsf{DySign} Family Attribution Evaluation Using Confusion Matrix
     }
   \label{fig:attribution_confmatrix}
\end{figure}
\end{scriptsize}

\textbf{Attribution Performance:} Figure \ref{fig:attribution_confmatrix} presents the confusion matrix for a more granular view of \textsf{DySign}'s family attribution.  The darker, in the matrix, is the diagonal, the more accurate is the attribution. However, due to the unbalanced malware families (Table \ref{tab:dataset_family_number}), there are some cells in the diagonal that are more darker because of the high number of samples in that family in the testing set. For this reason, we apply the log function on the original confusion matrix to have clearer results. Notice that all the produced results are based on the sandboxing reports of only $T=15s$ for each app whether it is a malware sample or a benign app. Therefore, the accuracy could be significantly improved by having a longer time $T$.

\begin{scriptsize}
\begin{figure}[ht!]
     \begin{center}
        \subfigure[Benign (Bytes)]{%
            \label{fig:ben_size_byte}
            \includegraphics[width=0.22\textwidth]{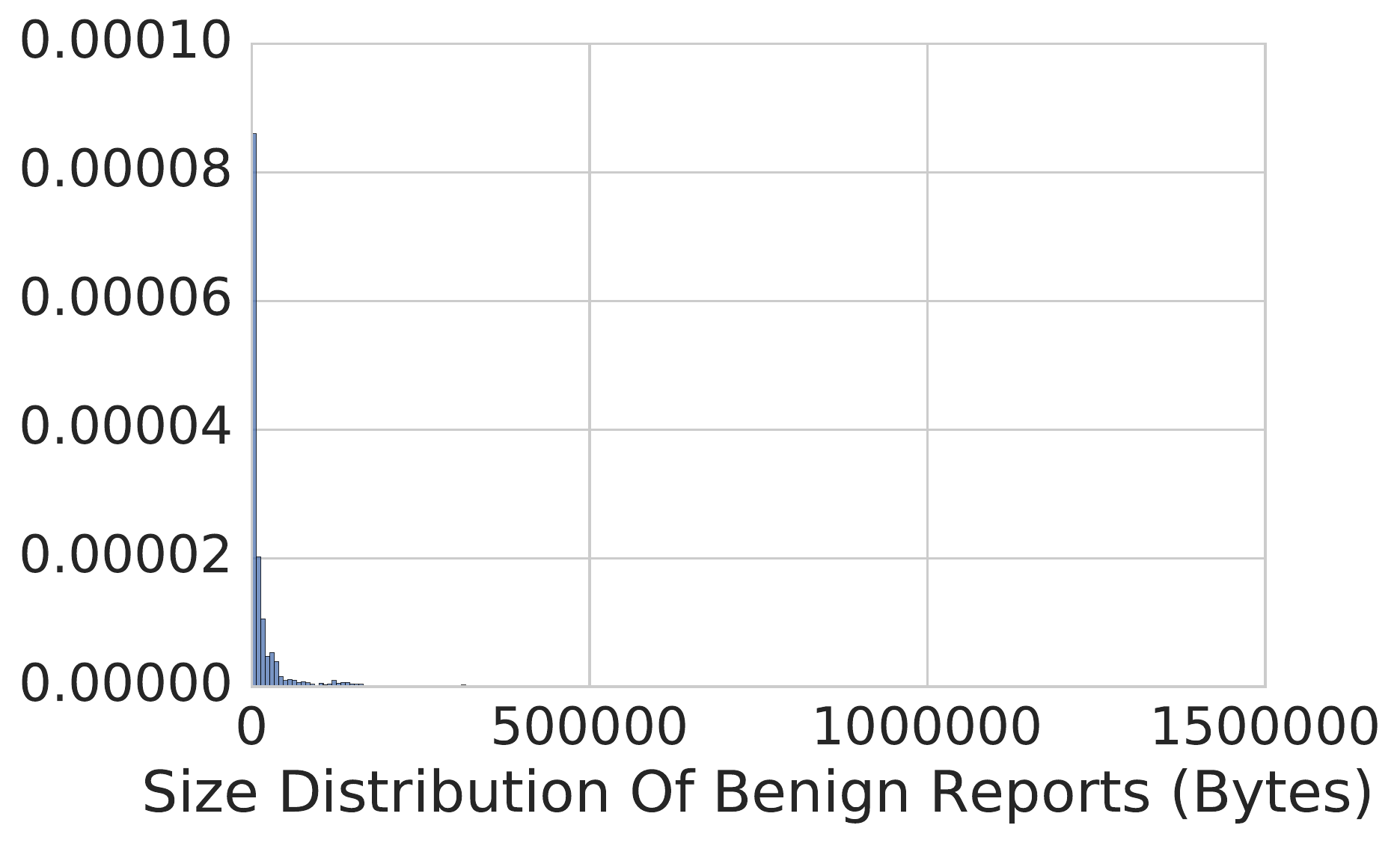}
        }
        \subfigure[Benign Log10 (Bytes)] {%
           \label{fig:ben_size_log}
           \includegraphics[width=0.22\textwidth]{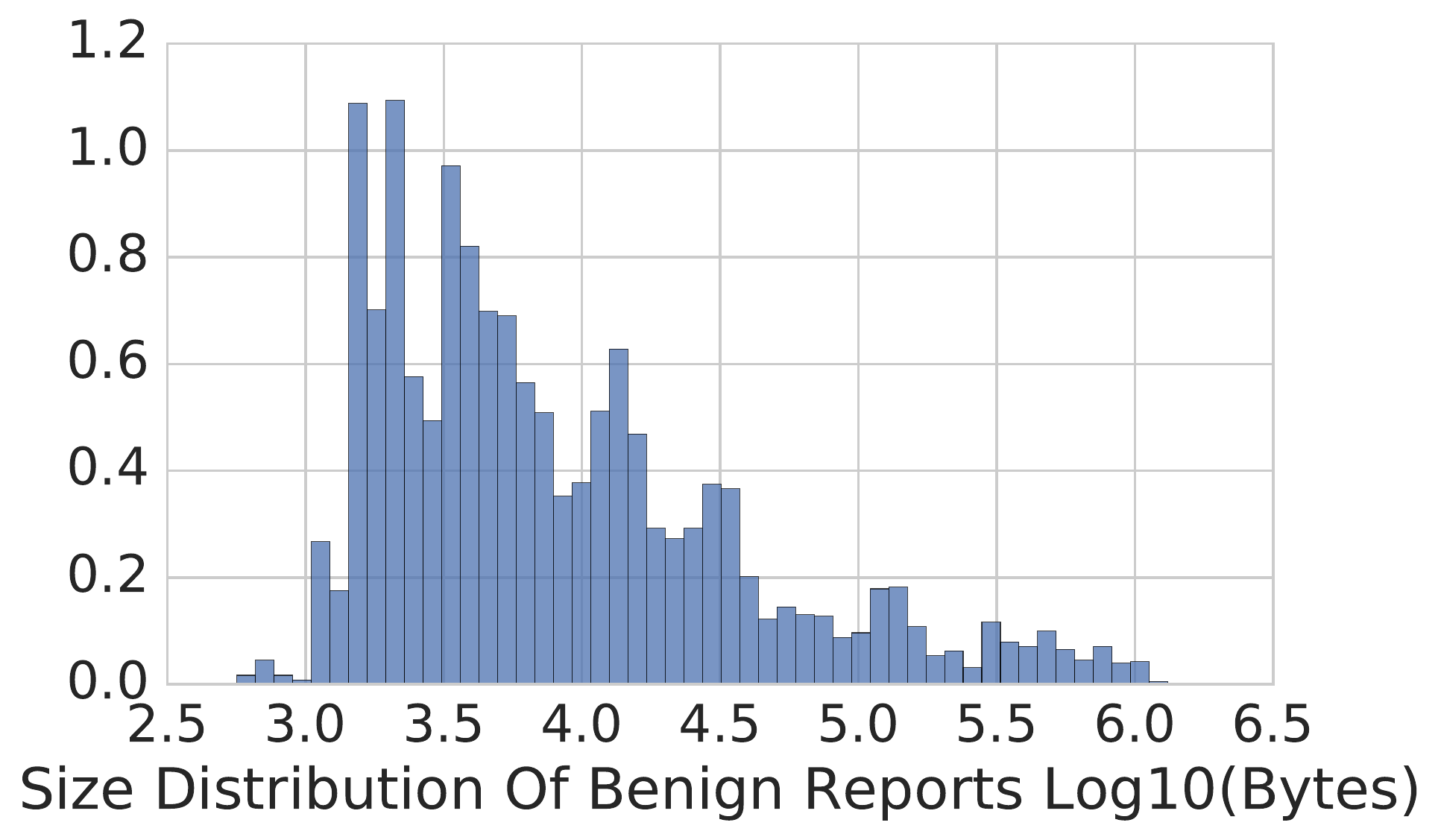}
        }
        \subfigure[Malware (Bytes)]{%
            \label{fig:mal_size_byte}
            \includegraphics[width=0.22\textwidth]{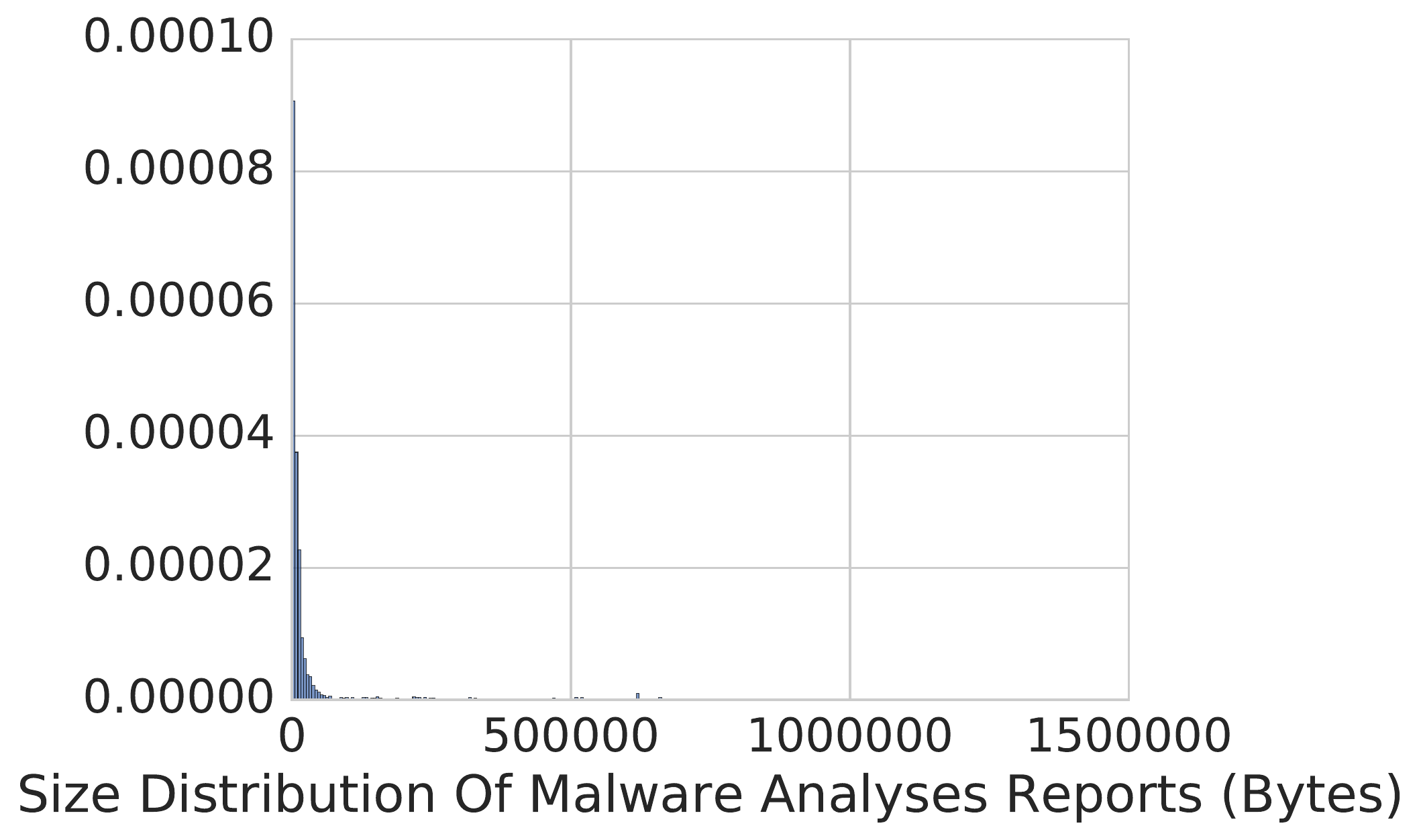}
        }
        \subfigure[Malware Log10 (Bytes)] {%
           \label{fig:mal_size_log}
           \includegraphics[width=0.22\textwidth]{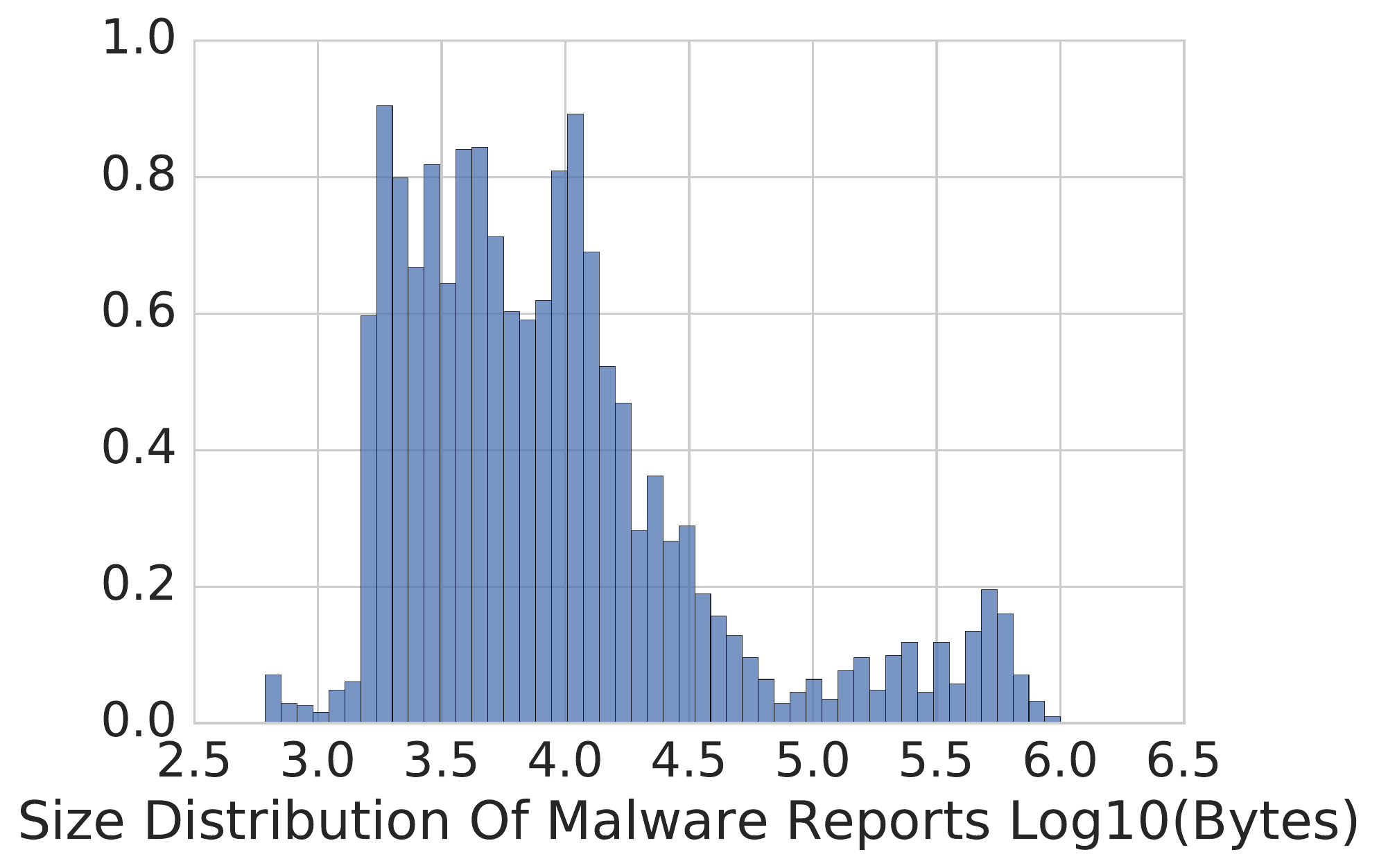}
        }
    \end{center}
    \caption{
        Sandboxing Output Size Distributions (Malware vs Benign)
     }
   \label{fig:size_ditribution_comparision}
\end{figure}
\end{scriptsize}

\textbf{Reports Size Analysis:} Figure \ref{fig:size_ditribution_comparision} shows the size distribution of the analysis reports. Figures \ref{fig:ben_size_byte} and \ref{fig:mal_size_byte} show the size distribution in \textit{bytes} for benign and malware reports respectively. To enhance the readability of the results, we apply the log function on the \textit{byte} distributions. The results are shown in Figures \ref{fig:ben_size_log} and \ref{fig:mal_size_log} for benign and malware reports. The most noticeable is the size of the malware comparing with benign reports. Malware reports tend to be bigger than benign ones. This difference happens in a very short time since we execute the apps for only $T=15s$. Our observations show that: i) Malicious apps tend to have similar behaviors and are generally eager to access the resources to perform their malicious tasks as soon as they are executed. ii) Malware apps tend to be self-driving, i.e., in most cases, they do not need UI interaction emulator. Instead, for example, they try to connect to a given IP address with a specific payload .

\begin{scriptsize}
\begin{figure*}[ht!]
     \begin{center}
        \subfigure[Detection F1-Score]{%
            \label{fig:ben_fscore}
            \includegraphics[width=0.31\textwidth]{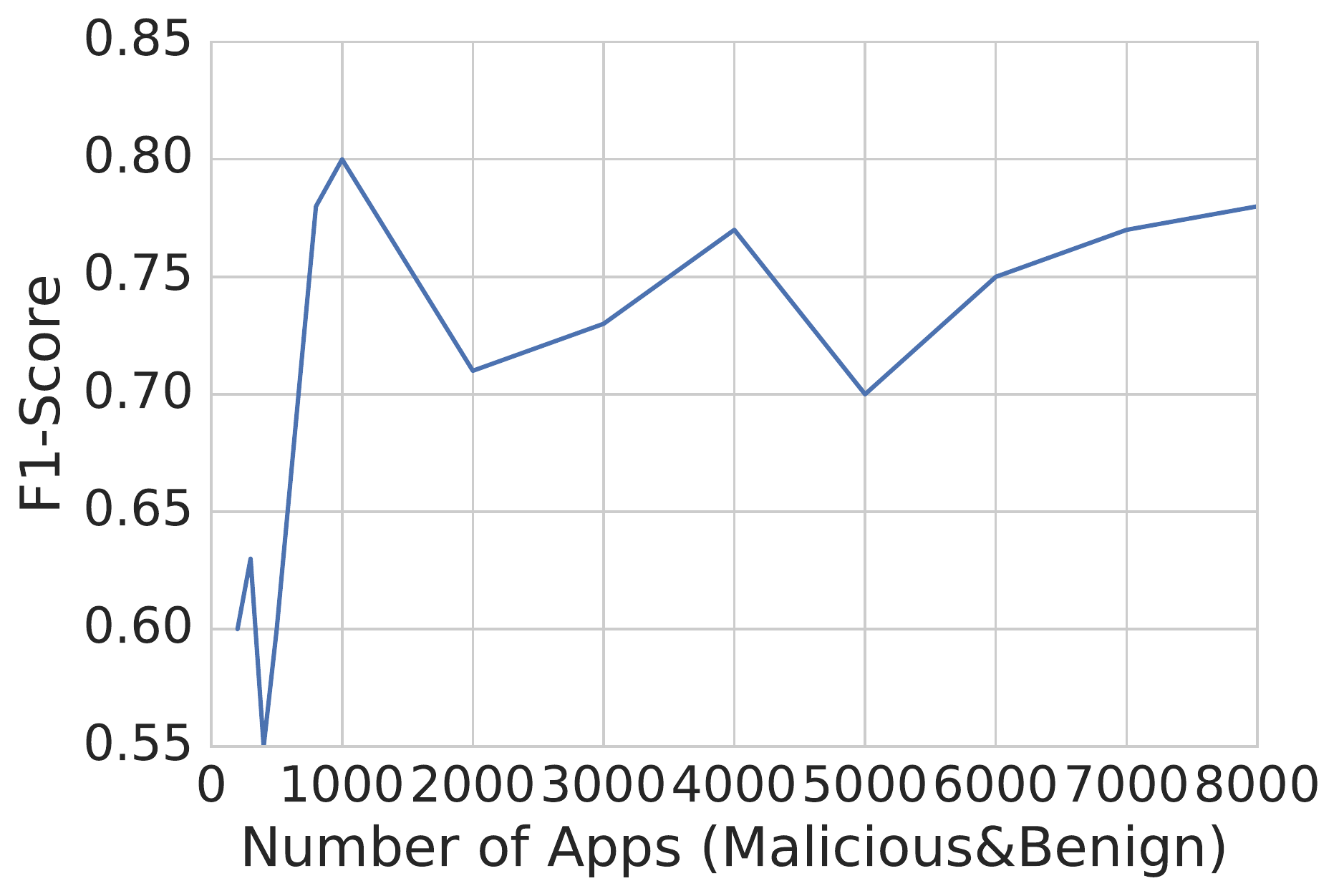}
        }
        \subfigure[Detection Precision] {%
           \label{fig:ben_precision}
           \includegraphics[width=0.31\textwidth]{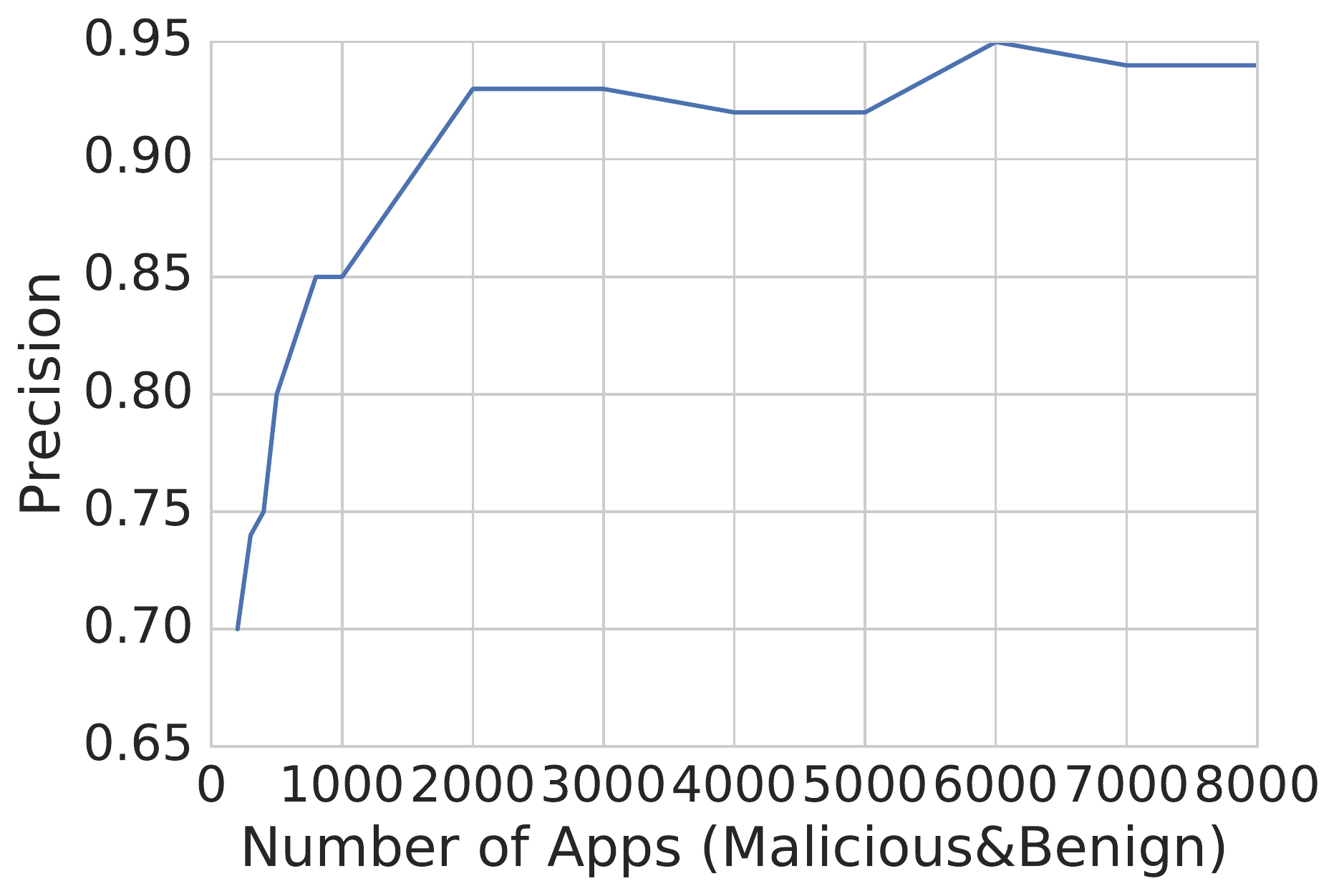}
        }
        \subfigure[Detection Recall]{%
            \label{fig:ben_recall}
            \includegraphics[width=0.31\textwidth]{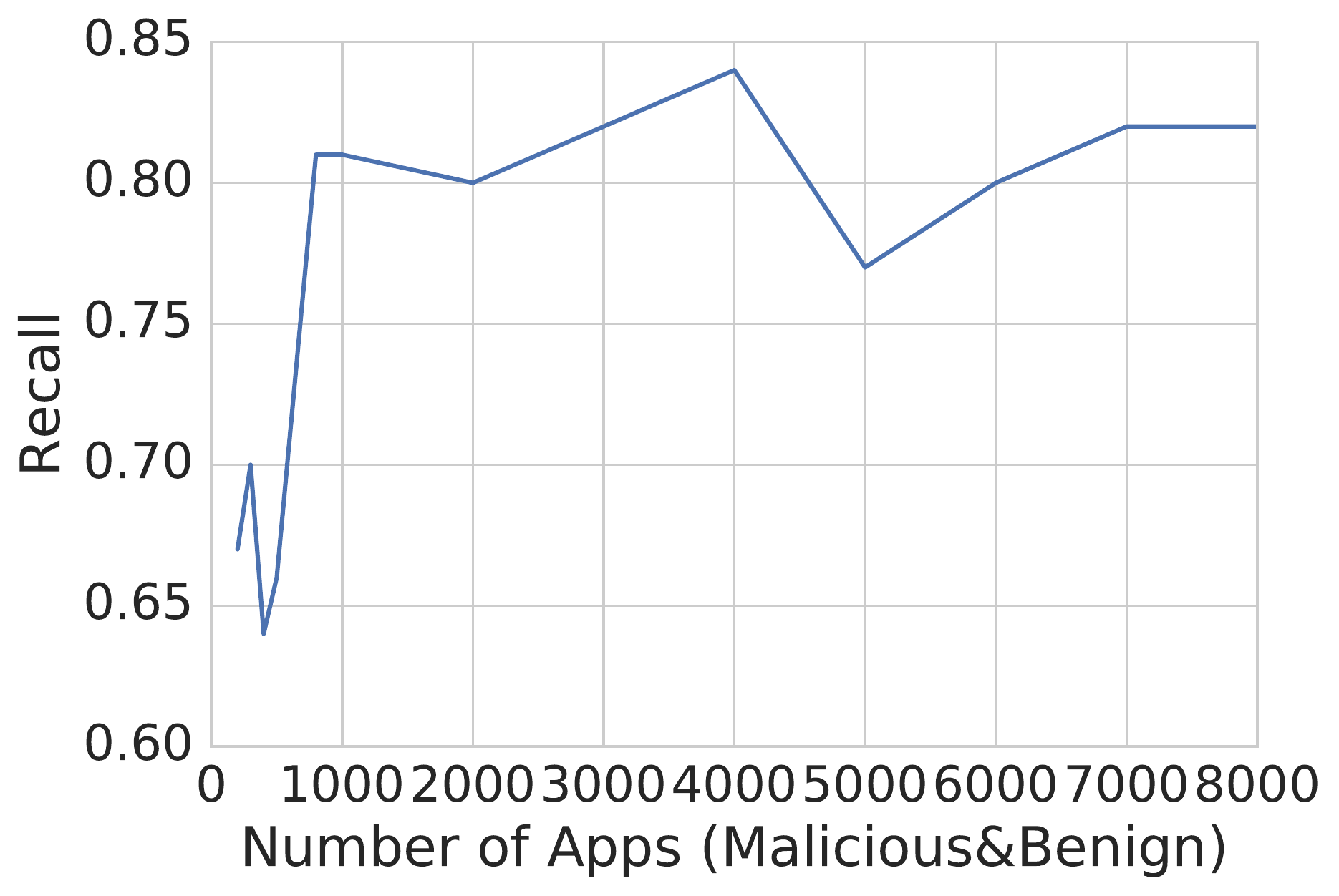}
        }
        
        \subfigure[Attribution F1-Score]{%
            \label{fig:ben_fscore}
            \includegraphics[width=0.31\textwidth]{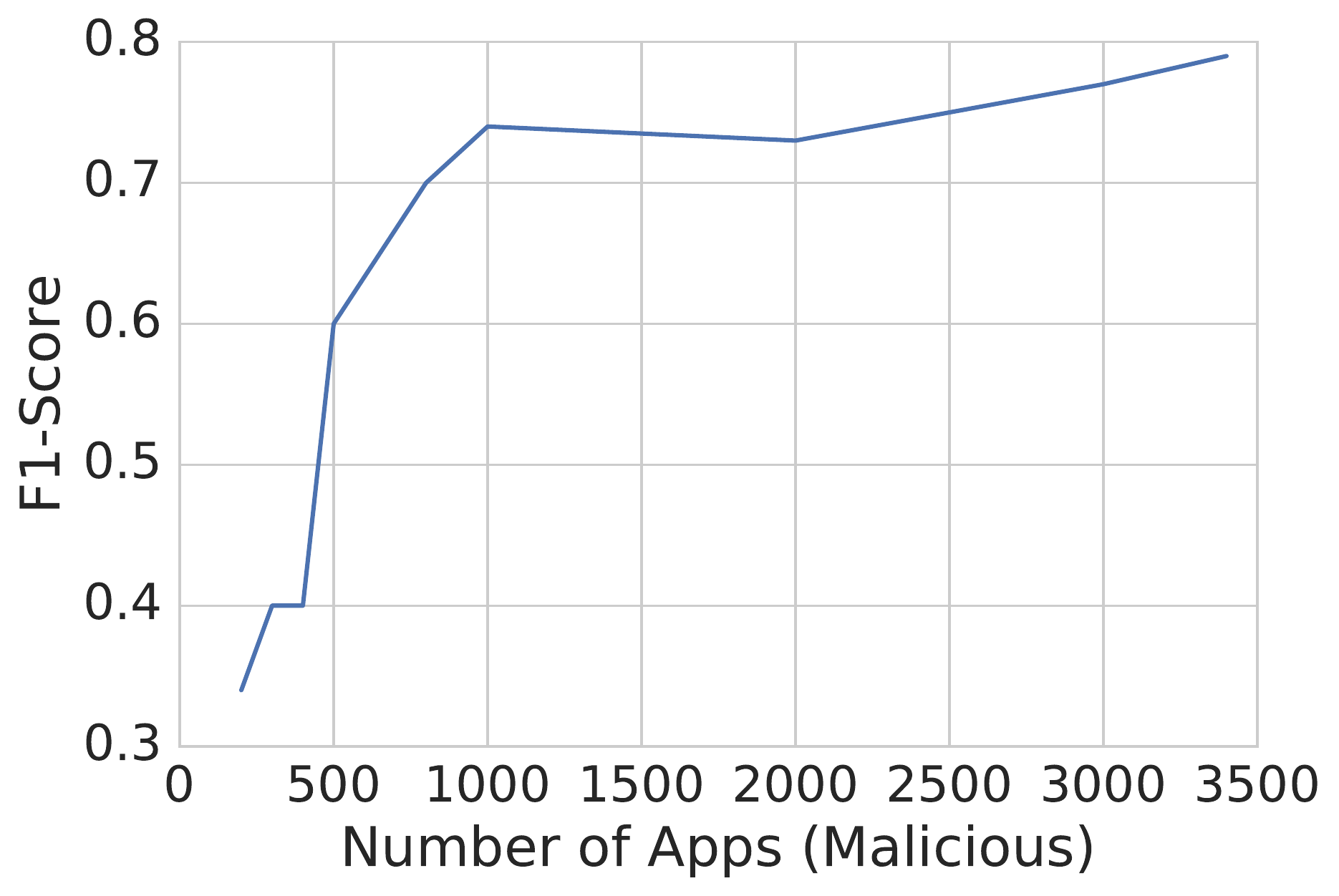}
        }
        \subfigure[Attribution Precision] {%
           \label{fig:ben_precision}
           \includegraphics[width=0.31\textwidth]{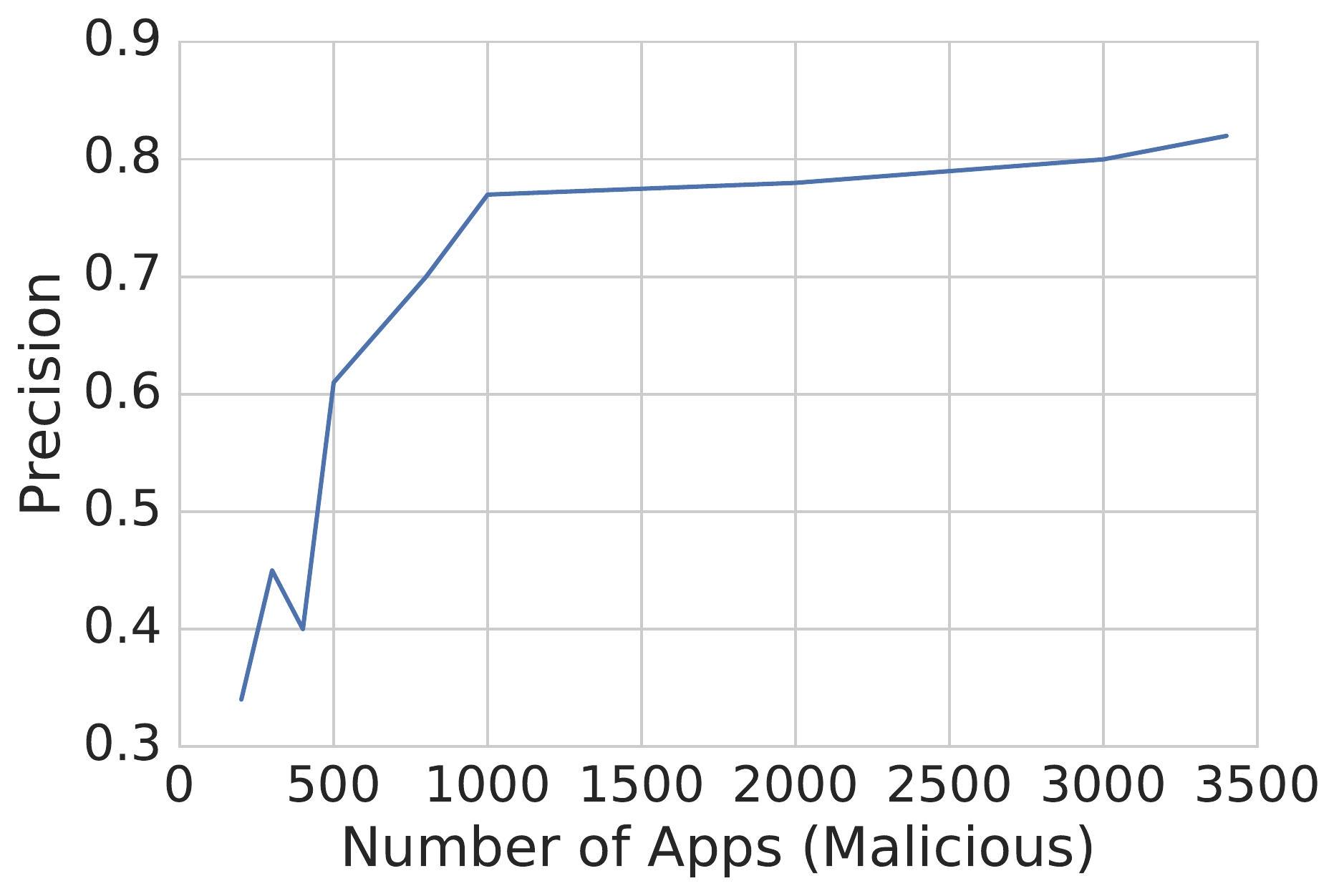}
        }
        \subfigure[Attribution Recall]{%
            \label{fig:ben_recall}
            \includegraphics[width=0.31\textwidth]{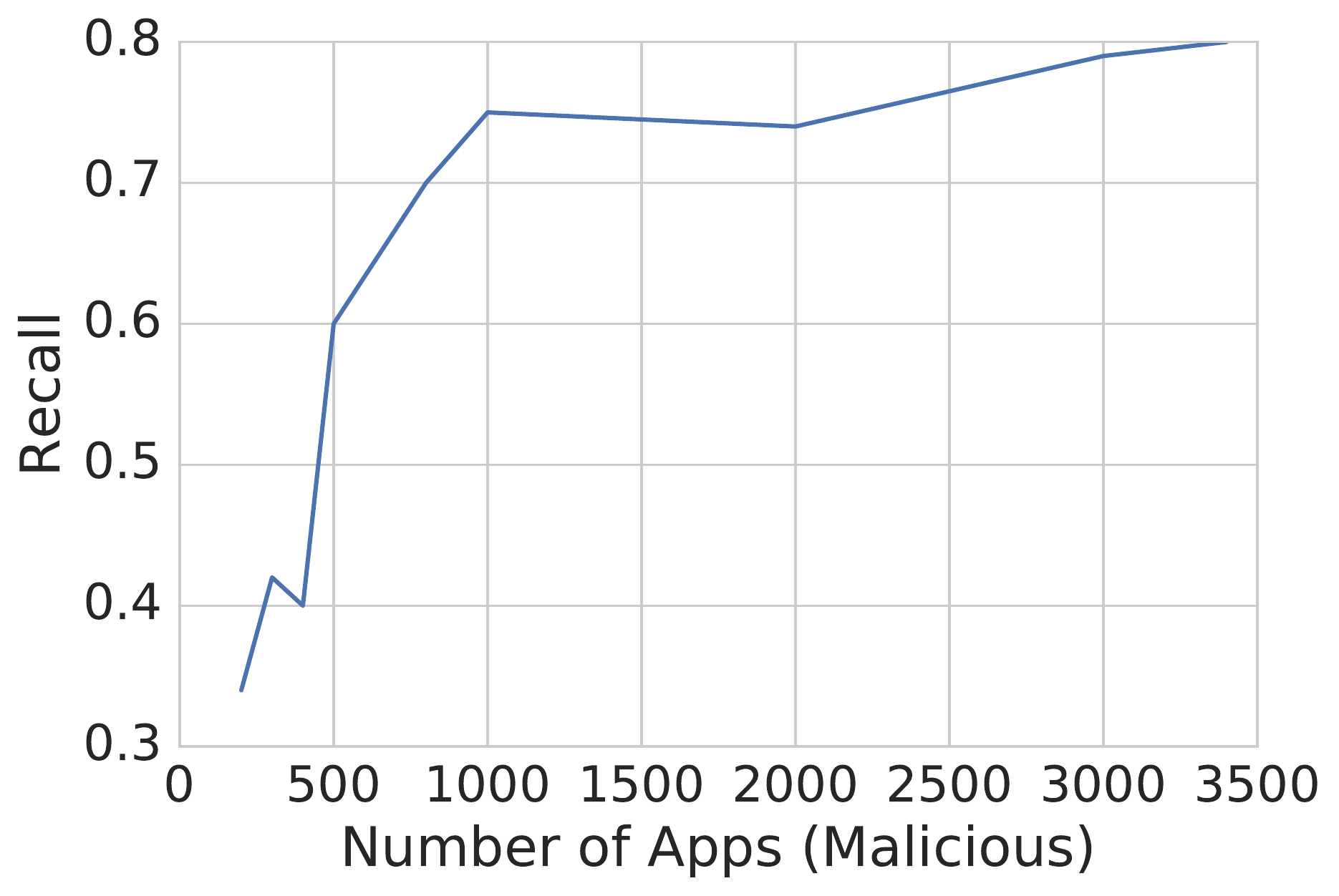}
        }
    \end{center}
    \caption{
        Detection and Family Attribution Performance Over Dataset Size
     }
   \label{fig:accuracy_scalability}
\end{figure*}
\end{scriptsize}

\textbf{Accuracy Performance and Dataset Size:} Figure \ref{fig:accuracy_scalability} shows the effect of the dataset size on the detection and family attribution. It also shows the direct relation between the number of samples in the dataset and the accuracy. The bigger is the dataset, the more accurate are the results. However, we could not test for higher scalability since we are limited by the size of Drebin dataset after excluding small families. According to the obtained results with our limited dataset, we conclude that by having a bigger dataset, \textsf{DySign} framework could achieve more accurate results. We let the validation of such conclusion as future work with much larger datasets.

\begin{scriptsize}
\begin{figure*}[ht!]
     \begin{center}
        \subfigure[Corpus Size vs Number of Reports]{%
            \label{fig:corpus_scalability}
            \includegraphics[width=0.3\textwidth]{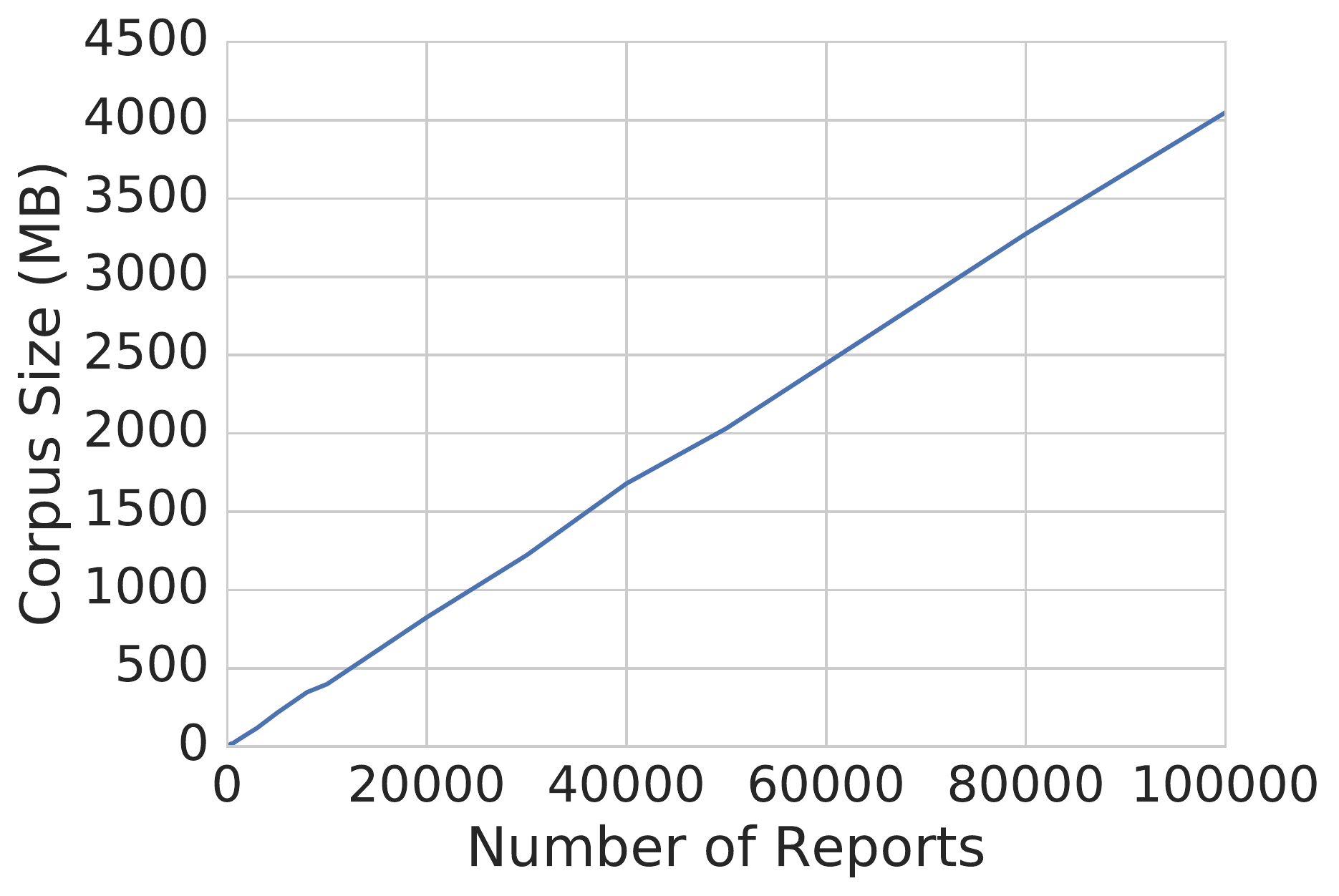}
        }
        \subfigure[LSH Matching Time vs Number of Reports] {%
           \label{fig:lsh_scalability}
           \includegraphics[width=0.3\textwidth]{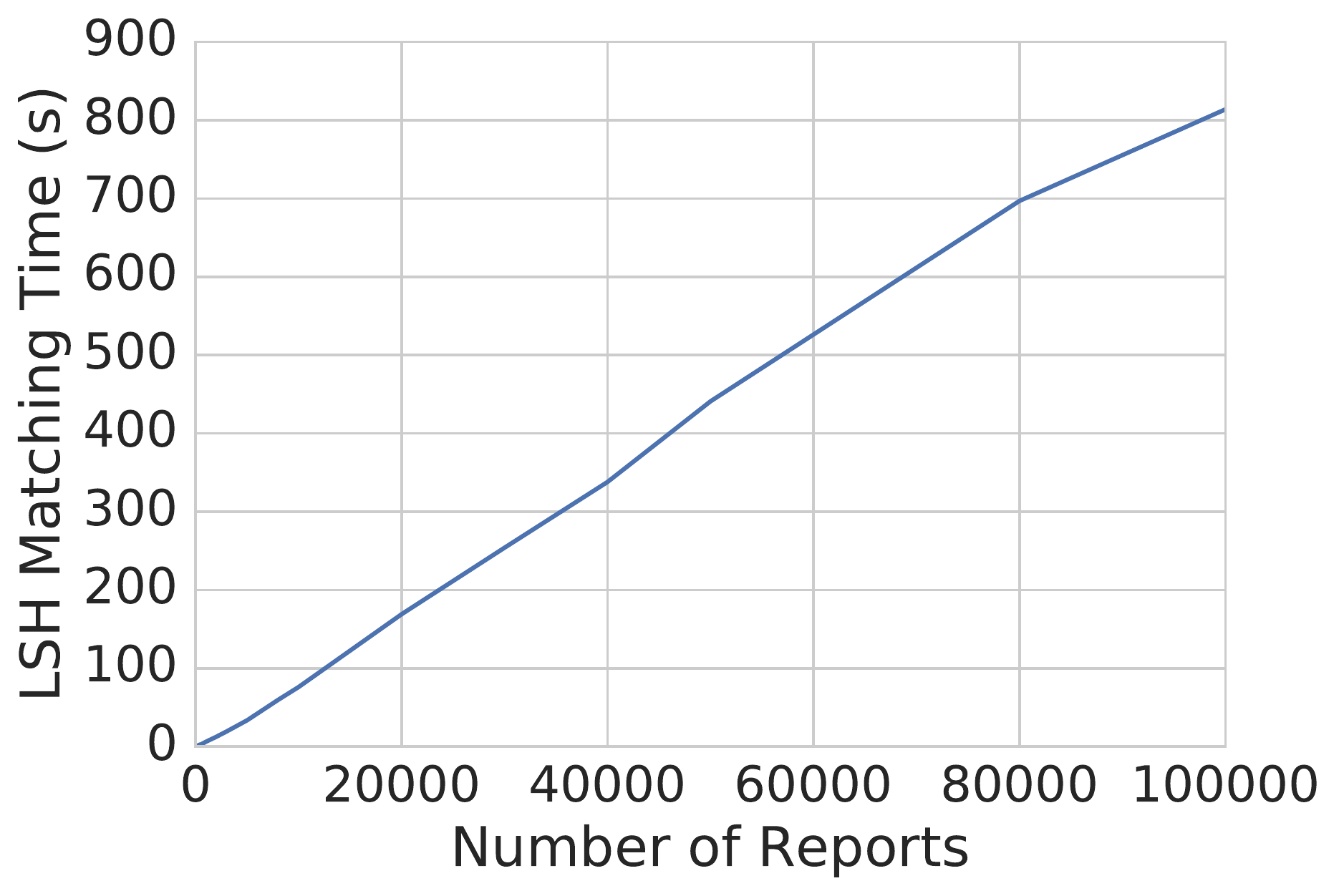}
        }
        \subfigure[TFIDF Computation vs Number of Reports]{%
            \label{fig:tfidf_scalability}
            \includegraphics[width=0.3\textwidth]{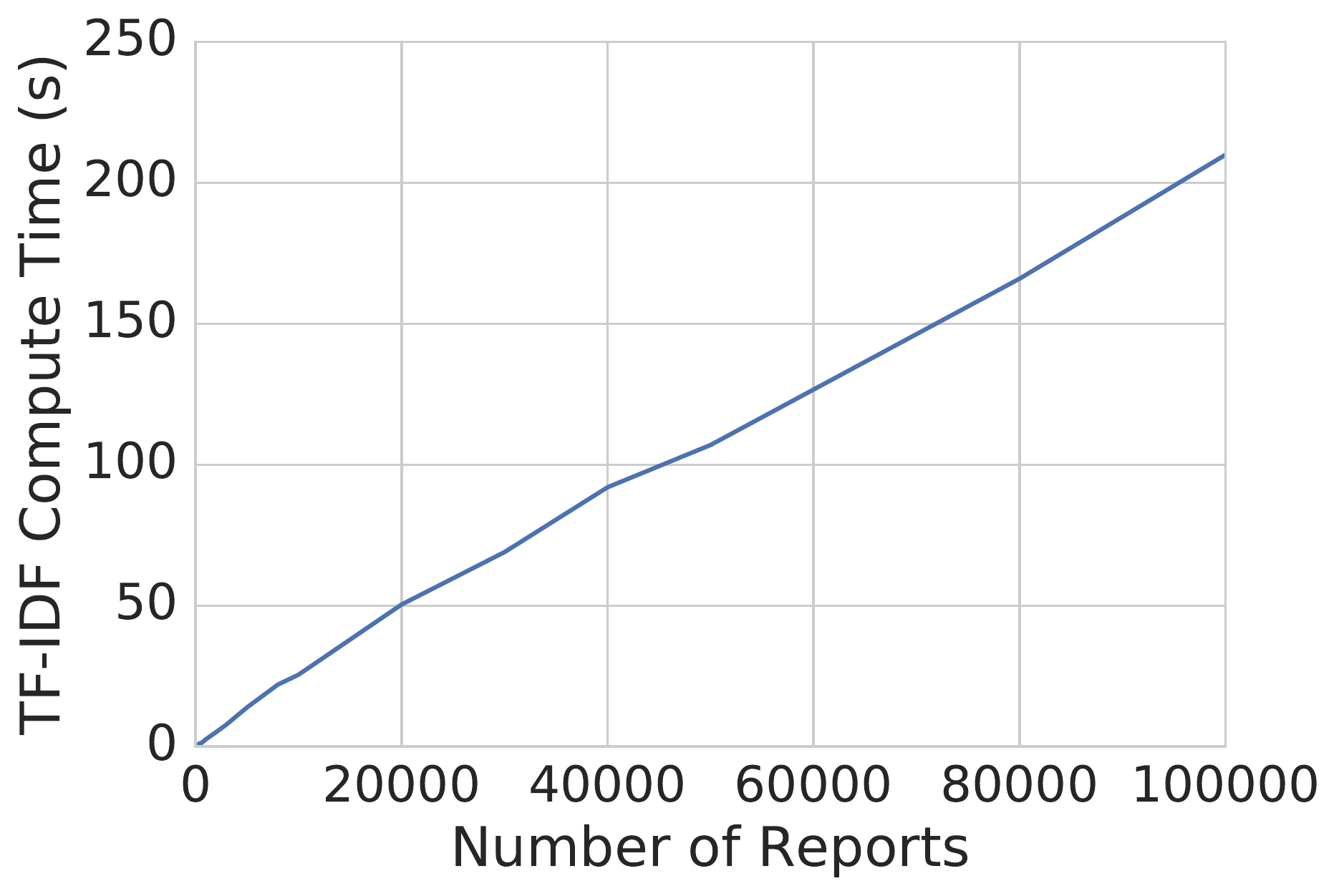}
        }
    \end{center}
    \caption{
        \textsf{DySign} Framework Scalability Analysis 
     }
   \label{fig:scalability_analysis}
\end{figure*}
\end{scriptsize}

\textbf{Scalability Analysis:} \textsf{DySign} shows high scalability, as summarized in Figure \ref{fig:scalability_analysis}. First, \textsf{DySign} computation time is very fast and linearly scalable with the number of reports. Our system could compute \textit{tf-idf} from $100,000$ analysis reports in about $200s$, as shown in Figure \ref{fig:tfidf_scalability}. Notice that we over-sample from our dataset in order to get  $100,000$ analysis reports used in the scalability evaluation. Figure \ref{fig:lsh_scalability} shows the linearity of \textit{LSH matching time} with the number of reports. Notice that for a $100,000$-report dataset, we match $10,000$ testing reports against  $90,000$ reports in the training dataset.

%===================================================

\section{Related Work} \label{sec:related_work}
In this section, we briefly introduce the existing works on Android malware analysis. They are categorized into static~\cite{arp2014drebin, feng2014apposcopy}, dynamic~ \cite{ali2016aspectdroid,canfora2016acquiring}, and hybrid~\cite{bhandari2015draco,zhang2014semantics}. 

\textbf{Static Analysis Approaches:} Static analysis techniques perform code disassembling and decompilation without actually running it. This approach is undermined by the use of various code transformation techniques~\cite{faruki2015android}. We may divide static analysis based techniques into the following categories: i) \textbf{Signature-based analysis:} This analysis deals with extracted syntactic pattern features~\cite{feng2014apposcopy, karbab2016apkdna, Karbab:2016:CBC:2991079.2991124}, and create a unique signature matching for a particular malware. However, such signature cannot handle new variants of existing known malware. Moreover, the signature database should be updated to handle new variants. AndroSimilar~\cite{faruki2013androsimilar} has been proposed to detect zero-day variants of the known malware. It is an automated  statistical feature signature-based method. However, it is sensitive due to code transformation methods. ii) \textbf{Resource-based analysis:} The Manifest file contains important meta-data about the components (i.e., activities, services, receivers, etc.) and required permissions.  There are some methods that have been proposed to extract such information and subject it to analysis~\cite{chin2011analyzing, fuchs2009scandroid}. iii) \textbf{Permission-based analysis:} Discovering the dangerous permission request is not adequate to proclaim a malware app, but nevertheless, permissions mapping requested and used permissions are an important risk identification technique~\cite{sarma2012android, barrera2010methodology}. %In~\cite{enck2009lightweight}, the authors proposed a certification tool that defines a set of rules to identify the combination of specific dangerous permissions to identify malware attributes. 

%iv) \textbf{Semantic-based analysis.} There are existing approaches that analyze Dalvik bytecode that is semantically rich containing type information such as classes, methods and instructions. Additionally, such information can be used to analyze control and data flow graphs that reveal privacy leakage and telephony services misuse\cite{grace2012riskranker, kim2012scandal}.

\textbf{Dynamic Analysis Approaches.} Dynamic analysis techniques  allow us to learn malicious activities. Android app execution is event-based with asynchronous multiple entry points. It is important to trigger those events.  
%One limitation of the dynamic approach is that some malicious execution paths may get missed if triggered according to some non-trivial events. Anti-emulation techniques such as Sandbox detection, timing out the analysis environment, and delaying malware execution can evade dynamic analysis methods.
 Dynamic techniques are divided into the following two categories: i) \textbf{Resources usage based:} Some malicious apps may cause Denial of Service (DoS) attacks by over-utilizing the constrained hardware resources. Range of parameters such as CPU usage, memory utilization statistics, network traffic pattern, battery usage and system-calls for benign and malware apps are gathered from the Android subsystem. Automatic analysis techniques along with machine learning techniques are used~\cite{shabtai2012andromaly,reina2013system, damopoulos2014best}. ii) \textbf{Malicious behavior based:} It is related to abnormal behaviors such as sensitive data leakage and sending SMS/emails~\cite{enck2014taintdroid,burguera2011crowdroid,elish2012user,huang2014asdroid}.

%===================================================
\section{Limitations and Concluding Remarks} \label{sec:conclusion}

We have reported, in this paper, the first investigation of the possibility of using dynamic features for  Android malware fingerprinting. {\sf DySign} leveraged state-of-the-art  \textit{machine learning} and \textit{Natural Language Processing} (NLP) techniques to produce agnostic fingerprints. The evaluation of   {\sf DySign} on both real-life malware and benign apps demonstrated a good detection and attribution performances with high scalability. Our work has a few limitations though. First, \textsf{DySign} fingerprinting approach is not deterministic, i.e., multiple executions could lead to slightly different fingerprints. However, the core information captured by such fingerprints is the same. Second, \textit{DySign} detection is limited by the Android malware families in  the analysis database, and therefore, it cannot detect malware belonging to new families. We plan to address these limitations in future work. In addition, we suggest exploring the applicability of a hybrid model in our detection system. 

%\section*{Acknowledgment}

\section*{Acknowledgements}
The authors would like to thank the anonymous reviewers for their insightful comments that allowed us to improve this paper.

\bibliographystyle{abbrv}
\bibliography{reference}

\end{document}